\documentclass[aps,superscriptaddress,pra,twocolumn]{revtex4-1}
\usepackage{amsmath}
\usepackage{graphicx}
\usepackage{dcolumn}
\usepackage{bm}
\usepackage[colorlinks,linkcolor=blue,anchorcolor=blue,citecolor=blue,urlcolor=blue,dvipdfm]{hyperref}

\begin{document}

\title{Bunching Effect and Quantum Statistics of Partially Indistinguishable
Photons}
\author{Fang-Wen Sun}
\email{fwsun@ustc.edu.cn}
\author{Ao Shen}
\author{Yang Dong}
\author{Xiang-Dong Chen}
\author{Guang-Can Guo}
\affiliation{CAS Key Lab of Quantum Information, University of Science and Technology of China, Hefei,
230026, P.R. China}
\affiliation{Synergetic Innovation Center of Quantum Information $\&$ Quantum Physics, University of Science
and Technology of China, Hefei, 230026, P.R. China}
\date{\today }

\begin{abstract}
The quantum statistics of particles is determined by both the spins and the indistinguishability of quantum states. Here we studied the
quantum statistics of partially distinguishable photons by defining the
multi-photon indistinguishability. The photon bunching coefficient was
formulated based on the properties of permutation symmetry, and a modified
Bose--Einstein statistics was presented with an indistinguishability induced
photon bunching effect. Moreover, the statistical transition of the photon state was
studied for partially distinguishable photons, and the results shows the
that indistinguishability exhibits the same role as that observed in the
generation of laser. The results will fill the gap between
Bose--Einstein and Poisson statistics for photons, and a formula is
presented for the study of multi-photon quantum information processes.
\end{abstract}

\maketitle

\section{Introduction}

The indistinguishability induced photon bunching effect is the foundation of
stimulated emission, multi-photon interference and general statistics \cite%
{QO,Sun07}. The stimulated emission process is the physical mechanism
underlying lasers and superluminescence. With multi-photon interference \cite%
{HOM,SA,PanRMP,Ou}, optical quantum information processing has been well
developed \cite{KLM,KokRMP,OBreinSci}, and the advantages have been
demonstrated in quantum computing via the Shor algorithm \cite%
{Lu07,Lanyon07,OBrien12}, boson sampling \cite%
{Broome,Spring,Tillmann,Bentivegna15}, and quantum metrology \cite%
{Vittorio11,Nagata07,SUNEPL,Xiang11}, which has achieved resolutions that
extend beyond classical limits and approach the quantum Heisenberg limit.
Additionally, for indistinguishable photons, the general photon number
distribution shows Bose--Einstein statistics. However, when photons are
partially distinguishable, the fidelity of quantum computing and the
resolution of quantum metrology quickly decreases as the photon numbers
increase, and in certain cases, the advantages of quantum information
processing can be lost. Moreover, the photon number distribution will vary
considerably from that of Bose--Einstein statistics. For example, the photon
number distribution could show Poisson statistics when the photons are
totally distinguishable. However, the properties of photon statistics have
not been clarified when photons are partially indistinguishable. Moreover,
the bunching effect of partially indistinguishable photons has not been
resolved.

In this study, we discuss the role of photon indistinguishability \cite%
{Sun09} in photon statistics. By defining and calculating the
indistinguishability ($K_{n}$) of an $n$-photon state, the photon bunching
effect is presented and analyzed in detail for partially indistinguishable
photons. Both the multi-photon indistinguishability and multi-photon
bunching effect show exponential decay with increase in the photon number.
Consequently, the photon statistical distribution is modified from the
Bose--Einstein statistics ($K_{n}=1$) by considering the partially
indistinguishable photon state and approaches the Poisson statistics when
the indistinguishability is lost ($K_{n}=0$). Because photon
indistinguishability induces notable photon bunching at high photon numbers,
the statistical transition of photon state may occur. Such a photon statistical
transition can be evaluated by the second-order degree of coherence where
the transition point highly depends on the photon indistinguishability.

In general, the statistical distribution of particles can be described as
\begin{equation}
P_{\varepsilon }\propto \frac{1}{\mathrm{e}^{\varepsilon /k_{B}T}-S}\text{,}
\end{equation}%
where $\varepsilon $ represents the energy, $k_{B}$ represents the Boltzmann
constant, and $T$ represents the absolute temperature. The statistical
properties of different particles are governed by their spins and the
indistinguishability of their quantum states. For indistinguishable Fermions
with half-integer spins, $P_{\varepsilon }$ represents Fermi--Dirac
statistics with $S=-1$; while for Bosons with integer spins, it shows
Bose--Einstein statistics with $S=1$. The main difference between these two
distributions is the value of $S$, which describes both the permutation
symmetric properties and the indistinguishability induced bunching factor.
%When the particles are indistinguishable with perfect permutation symmetry, $\left\vert S\right\vert =1$. However, when the particles are distinguishable, $S=0$. It indicates that the distinguishability is a crucial factor for the particle statistical.
In typical circumstances, particles are always interacting with other
particles or the outer environment, and their quantum coherence may be lost.
Thus, particles are in a mixed state and can be partially distinguishable.
In this case, the value of $\left\vert S\right\vert $ should be between $0$
and $1$. By studying the photon indistinguishability induced bunching effect
and photon statistics, we find that $S$ monotonously depends on the value of
the indistinguishability. This result will fill the gap in the photon
statistics between the indistinguishable case (Bose--Einstein statistics)
and the totally distinguishable case (Poisson statistics).

\section{Multi-photon indistinguishability and bunching effect}

Without a loss of generality, we consider a multi-photon state from $N$
separated emitters, which can be described as \cite{Sun09}
\begin{equation}
\text{\textsl{$\rho $}}_{NPhoton}=C_{0}\bigotimes_{k=1}^{N}(\left\vert \text{%
vac}\right\rangle \left\langle \text{vac}\right\vert +c_{k}\rho _{k})\text{,}
\label{single}
\end{equation}%
where $\rho _{k}$ ($\mathrm{tr}\rho _{k}=1$) describes the quantum state of
a single photon. $\left\vert \text{vac}\right\rangle $ is the vacuum state. $%
C_{0}$ is a normalization constant and $c_{k}>0$ is a constant determined by
the processes of photon generation and collection. For simplicity, we can
set all $c_{k}=c$ and $\rho _{k}=\rho $ because all emitters are under the
same environment during the photon generation process. A single photon might
be in a mixed state, which can be spectrally decomposed as $\rho
=\int_{-\infty }^{+\infty }\mathrm{d}\omega f(\omega )\left\vert \omega
\right\rangle \left\langle \omega \right\vert $ \cite{Sun09}, with $%
\left\vert \omega \right\rangle =\int_{-\infty }^{+\infty }\mathrm{d}%
\upsilon g_{\omega }(\upsilon )a^{\dag }(\upsilon )\left\vert \text{vac}%
\right\rangle $, where $a^{\dag }$ ($a$) is single photon creation
(annihilation) operator. $|g_{\omega }(\upsilon )|^{2}$ ($\int_{-\infty
}^{+\infty }|g_{\omega }(\upsilon )|^{2}\mathrm{d}\upsilon =1$) shows the
spectrum of the transform limited pulse with a center frequency $\omega $
and a width $\sigma _{g}$, and $f(\omega )$ ($\int_{-\infty }^{+\infty
}f(\omega )\mathrm{d}\omega =1$) is the distribution of a center frequency $%
\omega _{c}$ with a width $\sigma _{f}$.

To discuss the photon indistinguishability induced photon bunching effect
and photon statistics, we can define the indistinguishability of $n$ photons
as $K_{n}=\mathrm{tr}\rho ^{n}$, with $K_{2}\equiv K=\mathrm{tr}\rho ^{2}$
\cite{Sun09} and $K_{1}=\mathrm{tr}\rho =1$. Thus, when $\sigma _{f}=0$, the
single-photon state is a pure state and photons are indistinguishable with $%
K=1$ and $K_{n}=1$. However, because of interactions between single photons
and the outer environment or other photons in the generation process with $%
\sigma _{f}>0$, photons are partially distinguishable with $0<K_{n(n>1)}<1$.
When $\sigma _{f}\gg \sigma _{g}$, the photons are totally distinguishable,
with $K_{n}\longrightarrow 0$.

When both $g_{\omega }(\upsilon )=\mathrm{e}^{-(\upsilon -\omega
)^{2}/4\sigma _{g}^{2}}/\sqrt[4]{2\mathrm{\pi }\sigma _{g}^{2}}$ and $%
f(\omega )=\mathrm{e}^{-(\omega -\omega _{c})^{2}/2\sigma _{f}^{2}}/\sqrt{2%
\mathrm{\pi }\sigma _{f}^{2}}$ are Gaussian functions with widths of $\sigma
_{g}$ and $\sigma _{f}$, respectively, we can obtain $K=$ $\sigma _{g}/(%
\sqrt{\sigma _{g}^{2}+\sigma _{f}^{2}})$ \cite{Sun09}. In this case, $K_{n}$
can be analytically derived based on the value of $K$. Since
\begin{equation}
\left\langle \omega _{i}|\omega _{j}\right\rangle =\int\nolimits_{-\infty
}^{+\infty }g_{\omega _{i}}^{\ast }(\upsilon )g_{\omega _{j}}(\upsilon )%
\mathrm{d}\upsilon =\mathrm{e}^{-(\omega _{i}-\omega _{j})^{2}/8\sigma
_{g}^{2}}\text{,}
\end{equation}%
the value of $K_{n}$ can be
\begin{widetext}
\begin{eqnarray}
K_{n} &=&\int\nolimits_{-\infty }^{+\infty }\mathrm{d}\omega _{1}\mathrm{d}%
\omega _{2}\cdots \mathrm{d}\omega _{n}f(\omega _{1})f(\omega _{1})\cdots
f(\omega _{n})\left\langle \omega _{1}|\omega _{2}\right\rangle \left\langle
\omega _{2}|\omega _{3}\right\rangle \cdots \left\langle \omega _{n}|\omega
_{1}\right\rangle   \notag \\
&=&\frac{1}{(2\mathrm{\pi }\sigma _{f}^{2})^{n/2}}\int\nolimits_{-\infty
}^{+\infty }\mathrm{d}\omega _{1}\mathrm{d}\omega _{2}\cdots \mathrm{d}%
\omega _{n}\exp [\sum\nolimits_{i=1}^{n}(-\frac{\omega _{i}^{2}}{2\sigma
_{f}^{2}})]\times \exp [-\frac{(\omega _{1}-\omega _{2})^{2}+(\omega
_{2}-\omega _{3})^{2}+\cdots +(\omega _{n}-\omega _{1})^{2}}{8\sigma _{g}^{2}%
}]  \notag \\
&=&\frac{1}{(\sigma _{f}^{2})^{n/2}\sqrt{\det M_{n\times n}}}\text{,}
\end{eqnarray}%
\text{where the $n\times n$ matrix is}
\begin{equation*}
M_{n\times n}=\left[
\begin{array}{cccccc}
\frac{1}{2\sigma _{f}^{2}}+\frac{1}{4\sigma _{g}^{2}} & -\frac{1}{8\sigma
_{g}^{2}} & 0 & \cdots  & 0 & -\frac{1}{8\sigma _{g}^{2}} \\
-\frac{1}{8\sigma _{g}^{2}} & \frac{1}{2\sigma _{f}^{2}}+\frac{1}{4\sigma
_{g}^{2}} & -\frac{1}{8\sigma _{g}^{2}} & \cdots  & 0 & 0 \\
0 & -\frac{1}{8\sigma _{g}^{2}} & \frac{1}{2\sigma _{f}^{2}}+\frac{1}{%
4\sigma _{g}^{2}} & \cdots  & 0 & 0 \\
\vdots  & \vdots  & \vdots  & \ddots  & \vdots  & \vdots  \\
0 & 0 & 0 & \cdots  & \frac{1}{2\sigma _{f}^{2}}+\frac{1}{4\sigma _{g}^{2}}
& -\frac{1}{8\sigma _{g}^{2}} \\
-\frac{1}{8\sigma _{g}^{2}} & 0 & 0 & \cdots  & -\frac{1}{8\sigma _{g}^{2}}
& \frac{1}{2\sigma _{f}^{2}}+\frac{1}{4\sigma _{g}^{2}}%
\end{array}%
\right] \text{.}
\end{equation*}%
\end{widetext}In the above calculation, we simply set $\omega _{c}=0$ and
applied the $n$-dimensional Gaussian integral with
\begin{equation*}
\int\nolimits_{-\infty }^{+\infty }\exp (-\frac{1}{2}\sum%
\nolimits_{i,j=1}^{n}A_{i,j}x_{i}x_{j})\mathrm{d}^{n}x=\sqrt{\frac{(2\mathrm{%
\pi })^{n}}{\det A}}\text{,}
\end{equation*}%
where $A$ is a symmetric positive-definite $n\times n$ matrix \cite{GMatrix}%
. The first five terms are listed in Table.\ref{Tab}.

\begin{table}[t]
\caption{The values of multi-photon indistinguishability when spectral
distributions ($g_{\protect\omega }(\protect\upsilon )$ and $f(\protect%
\omega )$) of photons are Gaussian.}
\label{Tab}\centering\tabcolsep0.08in
\begin{tabular}{cccccc}
\hline\hline
$n$ & $2$ & $3$ & $4$ & $5$ & $6$ \\ \hline
$K_{n}$ & $K$ & $\frac{4K^{2}}{3+K^{2}}$ & $\frac{2K^{3}}{1+K^{2}}$ & $\frac{%
16K^{4}}{5+10K^{2}+K^{4}}$ & $\frac{16K^{5}}{3+10K^{2}+3K^{4}}$ \\
\hline\hline
\end{tabular}%
\end{table}

\begin{figure}[tbp]
\includegraphics[width=7cm]{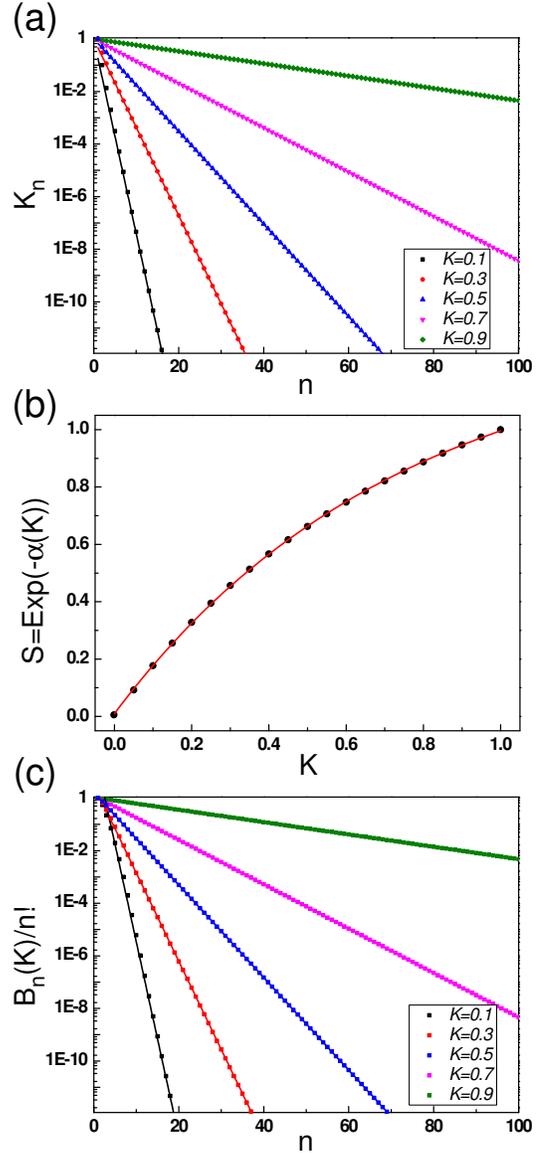}
\caption{(a) Multi-photon indistinguishability ($K_{n}$) shows exponential
decay with the photon number ($n$) for different two-photon
indistinguishabilities ($K$). The solid lines are the fittings with Eq.(%
\protect\ref{Kn}). (b) Photon bunching factor ($S=\mathrm{e}^{-\protect%
\alpha (K)}$) versus the two-photon indistinguishability ($K$). (c) Photon
bunching coefficient ($B_{n}$) with the photon number ($n$) for different
two-photon indistinguishabilities ($K$). The solid lines are the fittings
with Eq.(\protect\ref{BnK}). In the calculation, the spectral distributions (%
$g_{\protect\omega }(\protect\upsilon )$ and $f(\protect\omega )$) of the
photons are Gaussian. }
\label{Fig2}
\end{figure}

Fig. \ref{Fig2}(a) shows that the value of $K_{n}$ decays with an increase
in the photon numbers. We find that $K_{n}$ ($n\gg 1$) can be well fitted by
\begin{equation}
K_{n}(K)=\mathrm{e}^{-\alpha (K)n}  \label{Kn}
\end{equation}%
with a decay rate of $\alpha (K)$. Also, we can find that $%
K_{n+m}(K)=K_{n}(K)\times K_{m}(K)$. The value of $\mathrm{e}^{-\alpha (K)}$
is also shown in Fig.\ref{Fig2} (b). When $K=1$, $\mathrm{e}^{-\alpha (K)}=1$%
. Additionally, when $K\longrightarrow 0$, $\mathrm{e}^{-\alpha
(K)}\longrightarrow 0$.

Because the nonzero $K$ will induce photon bunching, the photon number
distribution of $\rho _{NPhoton}$ strongly depends on the value of $%
K_{n(n>1)}$. Formally, the photon state in Eq.(\ref{single}) can be
re-written as
\begin{equation}
\rho _{NPhoton}=C\sum_{n=0}^{N}\binom{N}{n}B_{n}c^{n}\{n\}\text{,}
\end{equation}%
where $C$ is a new normalization constant and $\{n\}$ describes the state
with the photon number of $n$. $B_{n}$ is an indistinguishability ($%
K_{n(n>1)}>0$) induced photon bunching coefficient.
\begin{figure}[tbp]
\includegraphics[width=7.5cm]{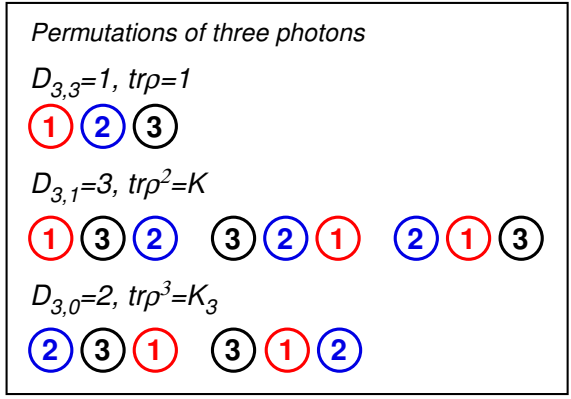}
\caption{Six ($3!=6$) permutations of three photons for the calculation of
three-photon indistinguishability.}
\label{Fig1}
\end{figure}

Principally, the Bosonic permutation symmetry induces the photon bunching
effect \cite{Sun07PRA}. Here we apply the permutation of $n$ photons to
obtain the photon bunching coefficient of an $n$-photon state, which can be
described as
\begin{equation}
B_{n}=\sum\limits_{k=2}^{n}D_{n,n-k}K_{k}+1\text{,}
\end{equation}%
where $D_{n,n-k}=\frac{n!}{(n-k)!}\sum\nolimits_{i=2}^{k}(-1)^{i}/i!$ are
rencontres numbers, which show the number of permutations of $n$ photons
with $(n-k)$ fixed photons without permutation. Fig.\ref{Fig1} illustrates
the number of permutations of three photons of $3!=6$, with $D_{3,3}=1$, $%
D_{3,2}=0$, $D_{3,1}=3$, and $D_{3,0}=2$. Thus, for totally distinguishable
states with $K=0$, when $n>1$, $K_{n}=0$ and $B_{n}=1$. For
indistinguishable states, $K_{n}=1$, $B_{n}=\sum%
\nolimits_{k=2}^{n}D_{n,n-k}+1=n!$ shows an $n$-photon bunching result and $%
\{n\}=\left\vert n\right\rangle \left\langle n\right\vert $ is an $n$-photon
Fock state. For partially indistinguishable photons, $1<B_{n}<n!$. When $%
n\gg 1$,
\begin{equation}
\frac{B_{n+1}(K)/(n+1)!}{B_{n}(K)/n!}\rightarrow \frac{K_{n+1}}{K_{n}}=%
\mathrm{e}^{-\alpha (K)}\text{,}  \label{Bn}
\end{equation}%
$B_{n}(K)/n!$ also shows an exponential decay with a photon number with a
decay rate of $\alpha (K)$. For the photon state with Gaussian spectral
distributions and $n\gg 1$,
\begin{equation}
B_{n}(K)=n!\mathrm{e}^{-\alpha (K)(n-1)}\text{,}  \label{BnK}
\end{equation}%
which is shown in Fig.\ref{Fig2} (c).

\section{Photon distribution of partially indistinguishable photons}

For totally distinguishable states, $B_{n}=1$, photon bunching does not
occur and $\rho _{NPhoton}$ shows a classical state with a binomial
distribution. When $N\gg 1$, the binomial distribution converts to Poisson
statistics \cite{OCQO}. For all indistinguishable states with $K_{n}=1$ and $%
B_{n}=n!$, the photon number distribution of Eq.(\ref{single}) is
\begin{equation}
\rho _{NPhoton}\simeq (1-Nc)\sum_{n=0}^{N}(Nc)^{n}\left\vert n\right\rangle
\left\langle n\right\vert =\sum_{n=0}^{N}P_{n}\left\vert n\right\rangle
\left\langle n\right\vert \text{,}
\end{equation}%
when $Nc<1$ and $N\gg 1$. It can be described by the Bose--Einstein
statistics with
\begin{equation}
P_{n}=\frac{\bar{n}^{n}}{(1+\bar{n})^{n+1}}=P\frac{\mathrm{e}^{-n\varepsilon
/k_{B}T}}{\mathrm{e}^{\varepsilon /k_{B}T}-1}\text{,}  \label{BE}
\end{equation}%
where $Nc=\mathrm{e}^{-\varepsilon /k_{B}T}$, $P=\mathrm{e}^{\varepsilon
/k_{B}T}+\mathrm{e}^{-\varepsilon /k_{B}T}-2$ and $\bar{n}=Nc/(1-Nc)=1/(%
\mathrm{e}^{\varepsilon /k_{B}T}-1)$ is the mean photon number.

However, for photons with partial indistinguishability ($0<K_{n}<1$), the
photon state should be
\begin{eqnarray}
\rho _{NPhoton} &\simeq &(1-Nc\mathrm{e}^{-\alpha (K)})\sum_{n=0}^{N}(Nc%
\mathrm{e}^{-\alpha (K)})^{n}\{n\}  \notag \\
&=&\sum_{n=0}^{N}P_{n}(K)\{n\} \text{.}  \label{Pn}
\end{eqnarray}%
When $Nc<1$ and $N\gg 1$, a modified Bose--Einstein statistics can be
presented as
\begin{equation}
P_{n}(K)=P(K)\frac{\mathrm{e}^{-n[\varepsilon /k_{B}T+\alpha (K)]}}{\mathrm{e%
}^{\varepsilon /k_{B}T}-S}\text{,}  \label{MBE}
\end{equation}%
where $P(K)=\mathrm{e}^{\varepsilon /k_{B}T}+\mathrm{e}^{-\varepsilon
/k_{B}T-2\alpha (K)}-2\mathrm{e}^{-\alpha (K)}$, and the mean photon number
is $\bar{n}=Nc\mathrm{e}^{-\alpha (K)}/(1-Nc\mathrm{e}^{-\alpha (K)})=1/(%
\mathrm{e}^{\varepsilon /k_{B}T+\alpha (K)}-1)$. Here, $S=\mathrm{e}%
^{-\alpha (K)}$ is an indistinguishability induced photon bunching factor.
Without changing $N$ and $c$, the statistics is modified from the
Bose--Einstein statistics in Eq.(\ref{BE}) via $S$, with $S=1$ for
indistinguishable case ($K=1$) and $S=0$ for the totally distinguishable
case ($K=0 $). The results clearly demonstrates the important role of
indistinguishability in photon statistics.

\section{Indistinguishability induced photon bunching and statistical transition}

\begin{figure}[tbp]
\includegraphics[width=7.5cm]{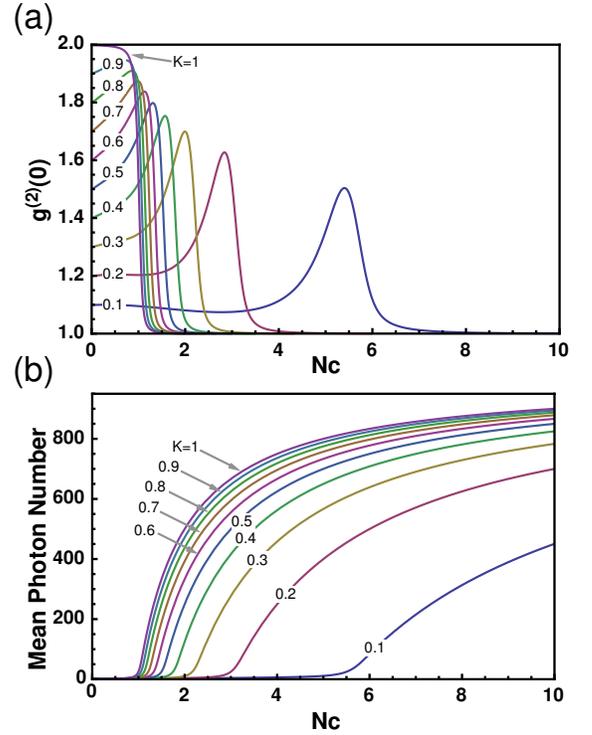}
\caption{(a) Second-order degree of coherence ($g^{(2)}(0)$) versus $Nc$.
(b) Mean photon number versus $Nc$. From left to right, the ten curves in
each figure correspond to two-photon indistinguishabilities ($K$) from $1$
to $0.1$. In the calculation, $N=1000$ and the spectral distributions ($g_{%
\protect\omega }(\protect\upsilon )$ and $f(\protect\omega )$) of the
photons are Gaussian.}
\label{Fig3}
\end{figure}

Because photons are Bosons, statistical transition can occur when more than one photon
occurs in a single mode, which results from the indistinguishability induced
photon bunching effect. Here, we apply the second-order degree of coherence (%
$g^{(2)}(0)$) to evaluate the photon statistical transition. For the
single photon state in Eq.(\ref{single}), $c$ describes the photon emission
probability from an emitter and $Nc$ is the number of photons from $N$
emitters without photon bunching. When $Nc\ll 1$, $g^{(2)}(0)=1+K$. However,
when $Nc\gg 1$ and $K>0$, more than one photon occurs in the emission mode
and the bunching effect from the indistinguishable multi-photon state
dominates the quantum statistics, as shown in Eq.(\ref{Bn}). This finding
demonstrates that photons condensate into an $n$-photon Fock state with $%
g^{(2)}(0)\rightarrow 1$ when $n\gg 1$. Fig.\ref{Fig3} (a) shows the values
of $g^{(2)}(0)$ with different values of $K$ and the behavior of photon
statistical transitions from $g^{(2)}(0)=1+K$ to $g^{(2)}(0)\rightarrow 1$
with an increase in the photon number $Nc$. For indistinguishable photon
state with Bose--Einstein statistics, the transition occurs at $Nc=1$. We
found that with lower $K$ values, a higher photon number $Nc$ is required to
make the transition. This finding indicates that photon indistinguishability
induced photon bunching effect is a key contribution to the transition. Eq.(%
\ref{Pn}) shows that, the transition points should occur approximately at $%
Nc=1/S$.

Such a transition can also be demonstrated by the mean photon numbers ($\bar{%
n}$) in Fig.\ref{Fig3} (b). At a low emission rate ($Nc\ll 1$), spontaneous
emission dominates and $\bar{n}$ increases slowly with $Nc$ for different
photon indistinguishabilities. However, when $Nc>1/S$, $\bar{n}$ increases
quickly with $Nc$ because the bunching effect from indistinguishable photons
induces stimulated emission \cite{Sun07} and dominates the photon
statistics. Higher $K$ values correspond to a higher increase rate. When $%
Nc\gg 1/S$, $\bar{n}\rightarrow N$, which demonstrates saturation. Fig.\ref%
{Fig3} shows that, although the photon emission in Eq.(\ref{single}) lacks
phase coherence, the photon indistinguishability exhibits the same role as
in the generation of laser \cite{Abmann}.

\section{Discussion and conclusion}

The multi-photon interference is essential for optical quantum information
processes. In addition to phase modulation, the photon indistinguishability
induced bunching effect is a key parameter in multi-photon interference.
Defining and calculating multi-photon indistinguishability ($K_{n}$) are key
elements in the analysis of multi-photon interference \cite%
{Sun07,Xiang06,RaNC} and optical quantum information processes. Because
multi-photon indistinguishability shows an exponential decay with increases
in the photon numbers,
%the indistinguishability of a single-photon source \cite{Somaschi,Ding} is a crucial requirement for optical quantum computation based on Knill-Laflamme-Milburn scheme \cite{KLM}.
such an imperfect indistinguishability is the reason for the exponential decay
in the fidelity of multi-photon entangled state and the visibility of
multi-photon interference \cite{Huang,Wang16}.
Especially in recently developed boson sampling \cite{Broome,Spring,Tillmann,Bentivegna15,Tillmann15} and quantum metrology \cite{Vittorio11,Nagata07,SUNEPL,Xiang11} with entangled photon number state, many photons interfere in a same spatial mode. Imperfect interference with partially indistinguishable photons highly decreases the fidelity of quantum computation and the resolution of the quantum metrology. Defining multi-photon indistinguishability provides important insights on these issues.

In conclusion, we have presented the definition of the indistinguishability
of multi-photon states. Based on the multi-photon emission model, we
discussed the indistinguishability induced bunching effect in photon
statistical behavior. The photon statistical distribution can be changed
from a classical Poisson distribution to Bose--Einstein statistics when the
multi-photon indistinguishability is increased from $0$ to $1$. A modified
Bose--Einstein statistics is presented for partially indistinguishable
photons with an indistinguishability induced photon bunching factor \cite%
{note}. In addition to its influence on photon statistical behavior,
multi-photon indistinguishability is a key parameter in multi-photon
interference for optical quantum information techniques and in the
generation of laser and superluminescence.

\section*{Acknowledgment}

This work is supported by the National Key Research and Development Program of China (No. 2017YFA0304504), the National Natural Science Foundation of China (Nos. 11374290,
91536219, 61522508, 11504363).


\begin{thebibliography}{99}
\bibitem{note} Although from Eq.(\ref{MBE}), the statistics of partially
indistinguishable photons can also be formally written as the Bose--Einstein
statistics with extra photon energy ($\varepsilon _{ex}=\alpha (K)k_{B}T$)
from the permutation of partially indistinguishable photons. However $%
K\rightarrow 0$, the extra photon energy is infinite.
\end{thebibliography}
\end{document}